# Anticipatory Governance in Data-Constrained Environments: A Predictive Simulation Framework for Digital Financial Inclusion

**Elizabeth Irenne Yuwono[a], Dian Tjondronegoro[a], Shawn Hunter[b], and Amber Marshall[a]**
[a]Department of Management, Griffith University, Brisbane, Australia, [b]Griffith Asia Institute, Griffith University, Brisbane, Australia

Corresponding author: e.yuwono@griffith.edu.au

**Abstract**

Financial exclusion presents a critical barrier to digital public service delivery, particularly in resource-constrained and archipelagic nations. Traditional policy evaluations often rely on retrospective descriptive data, failing to provide the *ex-ante* intelligence required for agile resource allocation. This study introduces a predictive simulation framework designed to support adaptive governance within government information systems. Leveraging the designated 2025 empirical baseline dataset from the UNCDF Pacific Digital Economy Program (N=10,108), we developed a three-stage methodological pipeline, comprising descriptive profiling, interpretable machine learning, and scenario simulation, to forecast the impact of digital financial literacy (DFL) interventions before deployment.

Utilizing cross-sectional structural associations, where behavioral levers function as constituent components of the literacy index, the framework projects intervention outcomes to serve as a prioritization heuristic. Prioritizing algorithmic transparency for public sector accountability, a Linear Regression model was selected ($R^2$=95.9%) to identify modifiable policy levers. Simulation results established a "Digital-First" sequencing protocol, demonstrating that foundational infrastructure interventions (e.g., device ownership) are projected to yield a 5.5% DFL improvement, significantly outperforming purely attitudinal nudges. The model enables precision targeting, identifying young female caregivers as a high-leverage segment while simultaneously flagging "non-responders" (urban professionals) to prevent fiscal waste. This research contributes to digital government scholarship by demonstrating how static survey data can be repurposed into actionable policy intelligence for anticipatory governance, offering a scalable, transparent, and evidence-based blueprint for bridging digital divides in data-fragmented environments. effects.

## 1. Introduction

Despite significant advancements in digital financial services (DFS) worldwide, financial exclusion remains a pervasive issue in many regions. Over the past decade, the expansion of access to these services has led to marked improvements in financial inclusion, with the proportion of adults holding accounts increasing globally from 51% in 2011 to 76% today (World Economic Forum, 2023). However, access alone is insufficient for effective utilization. Low levels of digital financial literacy (DFL) remain a critical barrier, preventing the translation of access into meaningful engagement with digital tools and, increasingly, with digital public services. DFL is defined as the key competencies for sound decision-making in acquiring and using DFS (Lyons & Kass-Hanna, 2021). Many segmented populations, particularly in low-income and rural communities, lack the knowledge to navigate these services safely, which directly affects their ability to engage with e-government platforms (Gallego-Losada et al., 2024; Mignamissi & Djijo T., 2021). As a result, governments have increasingly prioritized DFL education to address the digital financial divide.

However, a critical governance challenge remains: how to design evidence-based policy in data-constrained environments. The literature points to geographical dispersion as a key driver of exclusion, particularly in developing nations where infrastructure is fragmented. This challenge is particularly acute in the Pacific Island nations, where rural, low-income, and female demographics experience persistent exclusion due to the necessity of traversing vast ocean distances to access urban-centered services (UNCDF, 2023). In response, regional leaders endorsed the Money Pacific Goals 2025, identifying digital literacy as a high-priority target (Pacific Islands Forum Secretariat, 2024). Yet, existing literature confirms significant structural disparities across these nations, ranging from high proficiency in Fiji to deep exclusion in Papua New Guinea (PNG) (Tjondronegoro et al., 2025). Given these multidimensional challenges, an adaptive, data-driven approach is required to achieve inclusive growth.

To address this gap, the study introduces a predictive simulation framework designed to support evidence-based policymaking in resource-constrained environments. In this context, resource constraints refer to the intersection of informational constraints (specifically the lack of longitudinal or real-time administrative data), fiscal limitations that demand allocative efficiency, and the infrastructural fragmentation inherent to archipelagic geography (Gallego-Losada et al., 2024; Syauqi et al., 2021; Yan & Lyu, 2023). While this research utilizes the Pacific Island context as a compelling case study, the proposed methodological architecture suggests a framework adaptable to diverse low-resource contexts. By leveraging machine learning techniques to forecast the potential impact of targeted interventions on population-level DFL outcomes, the study presents a transferable approach for anticipatory governance that may be applicable to other data-fragmented regions, such as the Caribbean, Sub-Saharan Africa, and Southeast Asian archipelagos. This approach seeks to enable public sector stakeholders to design responsive, cost-effective strategies by identifying the most effective combinations of interventions for specific population groups.

Specifically, the study pursues the following objectives: (1) to profile the demographic, socio-economic, and behavioral characteristics of individuals with low and high levels of digital financial competency; (2) to employ a machine learning model to identify the most influential and modifiable factors associated with improvements in DFL; (3) to simulate the outcomes of targeted interventions across distinct capability domains, digital, traditional financial, and digital-financial; and (4) to evaluate the effectiveness and feasibility of the predictive simulation approach in informing forward-looking policy design. The simulations are framed as *ex-ante* projections derived from cross-sectional structural associations, where behavioral levers function as constituent components of the composite DFL index. While this design does not isolate temporal causality, it establishes a necessary prioritization heuristic for resource allocation in data-constrained environments .

By operationalizing machine learning within the policy cycle, this study contributes a scalable model for advancing digital equity. It demonstrates how governments can repurpose existing, periodic survey data to effectively target and bridge digital divides. This research provides insights for policymakers, development agencies, and financial institutions seeking effective, data-driven strategies to promote inclusive growth. By examining a comprehensive array of variables, it offers a nuanced understanding of why specific populations remain underbanked and how targeted, evidence-based interventions can effectively bridge persistent digital financial divides globally.

## 2. Theoretical Background

This section is organized into three main thematic areas that provide the essential context for the subsequent analysis. It first examines DFL, reframing it within the context of digital public service delivery and the structural factors driving exclusion. This section then moves to an assessment of predictive analytics and simulation, evaluating their role as decision-support tools in modern government information systems. Finally, it addresses the relevant theoretical foundations for adaptive policy design, anchoring the discussion within public governance concepts.

### 2.1. DFL as a Digital Public Service Challenge

Digital financial literacy has emerged as a foundational enabler not only of financial inclusion but of effective engagement with digital public services. As governments increasingly digitize welfare distribution, tax collection, and identification systems (Government-to-Person or G2P payments), DFL serves as a critical competency for citizens (Bhat et al., 2024; Choung et al., 2023). Defined as an integrated capability encompassing financial knowledge, digital competencies, and behavioral readiness, DFL equips individuals to effectively engage with the digital ecosystems, facilitating secure, informed, and autonomous participation in the digital economy and public administration (Bhat et al., 2024; Lyons & Kass-Hanna, 2021). In this context, DFL acts as a "digital bridge," facilitating the interaction between citizens and the state's digital infrastructure.

The significance of DFL is particularly evident in low-income and geographically isolated settings, where it serves as a mechanism to reduce administrative exclusion and buffer against digital risks such as fraud and misinformation (Mishra et al., 2024; Aziz and Naima, 2021). Consequently, strengthening DFL has become central to national strategies to enhance the resilience of digital public service delivery, especially for marginalized groups (Faulkner, 2022; Spivak et al., 2024). However, DFL remains unevenly distributed, creating a digital service divide that mirrors socio-economic and geographic inequalities. Empirical studies consistently highlight that low-income individuals, rural populations, and women exhibit significantly lower levels of DFL, attributable to limited digital infrastructure and socio-cultural constraints (Choudhary & Jain, 2023; Kass-Hanna et al., 2022).

These disparities are critical for policymakers because they result in uneven access to government resources. For instance, gender disparities in DFL are often exacerbated by the intersection of digital illiteracy and norms that limit women's access to the financial tools necessary to receive digital government aid (Mishra et al., 2024). Furthermore, a lack of trust in digital systems, concerns over security, and the perceived irrelevance of digital platforms hinder the uptake of e-government services (Han et al., 2023). Research indicates that DFL mediates this relationship, where higher digital financial knowledge enhances not just inclusion, but the behavioral capacity to navigate digital platforms securely, which is essential for user trust in digital governance. For instance, a study demonstrated that digital financial knowledge indirectly enhances financial inclusion through attitudinal and behavioral pathways, suggesting that affective and cognitive dimensions must be addressed jointly (Zaimovic et al., 2024). Similarly, another work underscored the importance of integrating DFL with broader constructs such as financial autonomy and capability, establishing it as a direct predictor of improved financial decision-making and subjective well-being (Kumar et al., 2023). These findings emphasize the need for multidimensional policy responses that align cognitive skills development with the emotional and contextual realities of citizens.

To address these gaps, governments often employ financial literacy training programs, which have shown some success in influencing behaviors such as savings and account ownership. However, the scalability and sustainability of these interventions remain limited, especially where digital components are underemphasized or poorly localized (Choudhary & Jain, 2023; Choung et al., 2023). This limitation is critical in multicultural nations, where unique infrastructural and cultural conditions necessitate highly contextualized program design (Hasan et al., 2021; Prasad et al., 2018). While recent studies have validated composite scales to capture key elements of DFL, such as fraud susceptibility, the literature lacks robust empirical models that simulate the dynamic interactions among individual characteristics, institutional frameworks, and intervention design (Lyons & Kass-Hanna, 2021; Ravikumar et al., 2022; Yadav & Banerji, 2023). For government information policies, this understanding underscores the imperative of adopting data-driven, context-sensitive frameworks that integrate behavioral insights to identify and target the specific DFL deficits preventing citizens from fully utilizing digital public services.

*2.2. Predictive Analytics and Simulation in Digital Governance*

The evolution of public policy for financial inclusion is increasingly shaped by the adoption of data-driven approaches that facilitate precision targeting and adaptive governance, a theoretical framework that emphasizes the capacity of institutions to learn, adjust, and evolve in response to complex, changing environments (Cleaver & Whaley, 2018; Karpouzoglou et al., 2016; van Assche et al., 2021). Within this framework, predictive modeling techniques emerge not merely as statistical exercises but as critical government information systems innovations (E. Tan et al., 2022; S. Y. Tan & Taeihagh, 2021). These tools function as decision-support mechanisms, enabling policymakers to transition from reactive measures to proactive refinement methods that anticipate community needs before they manifest as systemic exclusion. This capability is particularly vital in underserved and data-constrained environments, where the complexity of the digital divide demands high responsiveness from the government's digital delivery ecosystem.

However, a significant gap exists in the current government information systems landscape regarding the maturity of analytical capabilities. The existing literature distinguishes between descriptive analytics, which dominate current public sector dashboards and focus on retrospective reporting (e.g.,

"what happened?"), and predictive analytics, which offer anticipatory intelligence (e.g., "what will happen?") (Xin et al., 2024; Zheng & Shuang, 2025). Despite the proliferation of inclusion programs globally, the integration of predictive analytics into DFL interventions remains limited. Existing policy evaluations primarily rely on retrospective methods, such as before-and-after surveys or cross-sectional trend analysis (Levantesi & Zacchia, 2021; Zhu, 2025). This omission constrains the ability of public administrators to simulate the "learning" and "adaptation" processes central to adaptive governance, limiting policy dashboards to static reporting rather than dynamic forecasting.

Where applied, predictive modeling has demonstrated its potential to enhance the intelligence of digital government platforms. In the Italian context, Levantesi and Zacchia (2021) employed ensemble machine learning techniques to uncover nonlinear determinants of financial knowledge, enabling the segmentation of populations into targeted support categories. Similarly, Zhu (2025) applied advanced algorithms to youth financial literacy in Hong Kong, demonstrating that integrating interdisciplinary features substantially improves the predictive accuracy of policy models. These studies affirm that machine learning can provide the granular understanding of population subgroups necessary to inform tailored government interventions.

Beyond individual-level prediction, the relevance of predictive analytics extends to system-level policy planning and digital ecosystem management. Studies advocate for AI-based models that align service personalization with unique financial profiles and use predictive analytics to identify regional disparities for inclusive growth (Akanfe et al., 2025; Xin et al., 2024). These approaches underscore the capacity of machine learning models to serve as anticipatory governance tools, allowing officials to forecast the effects of future programs, an essential capability for resource-constrained governments seeking to optimize their digital portfolios.

To implement this within a policy framework, this study employs static scenario modelling (Cairns et al., 2004). It is important to distinguish this approach from complex Agent-Based Modeling (ABM) or System Dynamics, which require massive, real-time longitudinal data streams often unavailable in developing nations (Axtell & Farmer, 2025; Chang et al., 2022; Sukhwal & Kankanhalli, 2022). While ABM offers the capacity to simulate emergent behaviors and temporal evolution within a system, its application is often constrained by the requirement for high-frequency longitudinal data streams, which are frequently unavailable in developing nations. In contrast, static scenario modeling is more suitable for data-constrained environments because it leverages accessible cross-sectional data to generate predictive insights without necessitating the continuous longitudinal streams required by dynamic models. Although this approach does not capture dynamic feedback loops or time-dependent adaptation, it provides a rigorous, transparent heuristic for estimating intervention impact where full-scale dynamic simulation is not viable. For data-constrained and developing regions, where programmatic experimentation is costly, this form of scenario modeling reduces the risk of policy failure.

Moreover, existing studies largely lack region-specific adaptation, a key requirement for effective information systems. Predictive models developed in high-resource or urban contexts often do not generalize well to the sociocultural and infrastructural conditions of low-resource or fragmented geographies (Luo et al., 2022; Yu, 2022). Adaptation of such models requires the incorporation of localized behavioral data and institutional trust indicators to ensure relevance.

In summary, predictive modeling offers a powerful yet underexploited opportunity to enhance digital governance. By revealing hidden patterns and forecasting outcomes, these tools support evidence-based decision-making. While global examples illustrate the potential of predictive approaches, their integration into government information systems remains nascent. Bridging this gap necessitates the development of simulation-based policy frameworks that support adaptive governance through robust, transparent, and predictive analytics.

*2.3. Theoretical Foundations for Adaptive Policy Design*

To connect predictive analytics within the public sector, this study anchors its approach in adaptive governance, a theoretical framework suited for managing complex social-ecological systems characterized by high uncertainty and structural inequality (Karpouzoglou et al., 2016; Nookala, 2024). In the context of archipelagic and resource-constrained developing nations, where geographic isolation and systemic capacity gaps prevail, static policy models are insufficient. Adaptive governance posits that institutions must be capable of iterative learning and rapid adjustment to local feedback. Within this study, the predictive simulation tool functions as the technological enabler of this theory, providing the computational feedback loops necessary for policymakers to anticipate outcomes and adjust DFL strategies before implementation.

This shift toward adaptive, data-driven policymaking intersects with the concept of Digital Bureaucracy and the framework of collective ambidexterity (Hietala & Päivärinta, 2025; Peeters, 2023). The framework highlights the tensions involved in managing the balance between centralized control and localized innovation in digital service delivery. It situates financial inclusion within the broader digital governance models, underscoring the importance of coherent coordination across policy levels, central authorities, community institutions, and user groups. For geographically fragmented states, where decentralization is a structural necessity, the predictive model serves as a centralized policy dashboard that informs localized action, resolving the tension between national standardization and community-specific needs.

Public policy scholarship further emphasizes that digital service delivery must be embedded within sustainable governance systems. A recent study proposed a framework situating digital public services within such systems, underscoring the role of targeted literacy interventions in building trust among marginalized communities (Djatmiko et al., 2025). These insights echo findings from development initiatives in similar contexts, where trust in institutions is critical to the uptake of digital services (Bharathi. S et al., 2023). Furthermore, another work argued that government disclosure and transparency are drivers of institutional trust (Ripamonti, 2024). By employing explainable simulation models, governments can enhance transparency regarding how and why specific interventions are targeted, thereby fostering greater public trust in the digital state.

Behavioral theories of public administration provide the final layer of this theoretical foundation. Financial literacy is argued to be the mediating factor between public policy and inclusion, advocating for designs that account for the behavioral dimensions of decision-making (Khan et al., 2022). These propositions align with the need for behaviorally informed government information systems that can ingest data on cultural norms, such as communal decision-making, and output policy scenarios that respect these realities. Similarly, another study offered a taxonomical framework for digital vulnerability, enabling policy actors to classify and target marginalized groups (Pérez-Escolar & Canet, 2023). The predictive simulation framework supports this taxonomy, transforming static vulnerability categories into dynamic variables for scenario testing.

However, a critical gap remains in the literature. While frameworks like collective ambidexterity, sustainable governance, and digital vulnerability describe what sustainable inclusion looks like, they often lack the analytical instruments to support these concepts in resource-constrained environments. Most theoretical models originate in high-capacity contexts and do not account for the data fragmentation typical of developing administrative contexts. Consequently, their direct transferability is limited without recalibration. This study addresses this theoretical gap by introducing a predictive simulation framework as the necessary mechanism to translate these abstract governance theories into an actionable, evidence-based strategy. By leveraging machine learning, the study demonstrates how adaptive governance can be translated as a computational mechanism, moving from theoretical ideals to practical, data-driven public administration.

*2.4. Gaps and Contributions*

Despite the imperative for adaptive governance in digital public service delivery, a significant disconnect remains between theoretical aspirations and reality. While existing literature highlights the structural drivers of the digital service divide, particularly among rural, low-income, and female populations, policy responses often rely on static, retrospective evaluations. These conventional approaches overlook the complex behavioral dynamics and non-linear interactions characteristic of low-capacity, archipelagic nations. In such environments, where geographic fragmentation and limited state presence demand high responsiveness, the absence of predictive tools renders true adaptive governance theoretically attractive but operationally elusive.

Theoretical frameworks for public policy, as discussed in the previous section, provide the normative architecture for inclusion but often lack the computational mechanisms required for implementation. Concepts such as digital vulnerability and collective ambidexterity explain why adaptation is necessary, but they do not offer the decision-support instruments needed to simulate how specific interventions will perform before deployment. Furthermore, while machine learning has transformed private sector fintech, its application within government information systems for social policy design remains rare.

Most public sector studies fail to simulate intervention outcomes or model the heterogeneous effects of policy shifts on underserved subgroups, limiting the ability of governments to design scalable, equity-driven strategies.

This study directly addresses these conceptual and methodological gaps. Its primary theoretical contribution is the provision of a predictive simulation framework that supports adaptive governance within a resource-constrained context. Fig. 1 illustrates the conceptual model. By leveraging machine learning to model modifiable behavioral drivers, the study provides the necessary analytical core to transform static policy theory into anticipatory policymaking. This innovation allows decision-makers to conduct *ex-ante* evaluations, modeling the population-level change of targeted interventions on DFL prior to resource commitment. Ultimately, this research offers a replicable blueprint for data-enabled governance, demonstrating how predictive analytics can bridge the gap between abstract policy goals and the tangible realities of the digital financial divide.

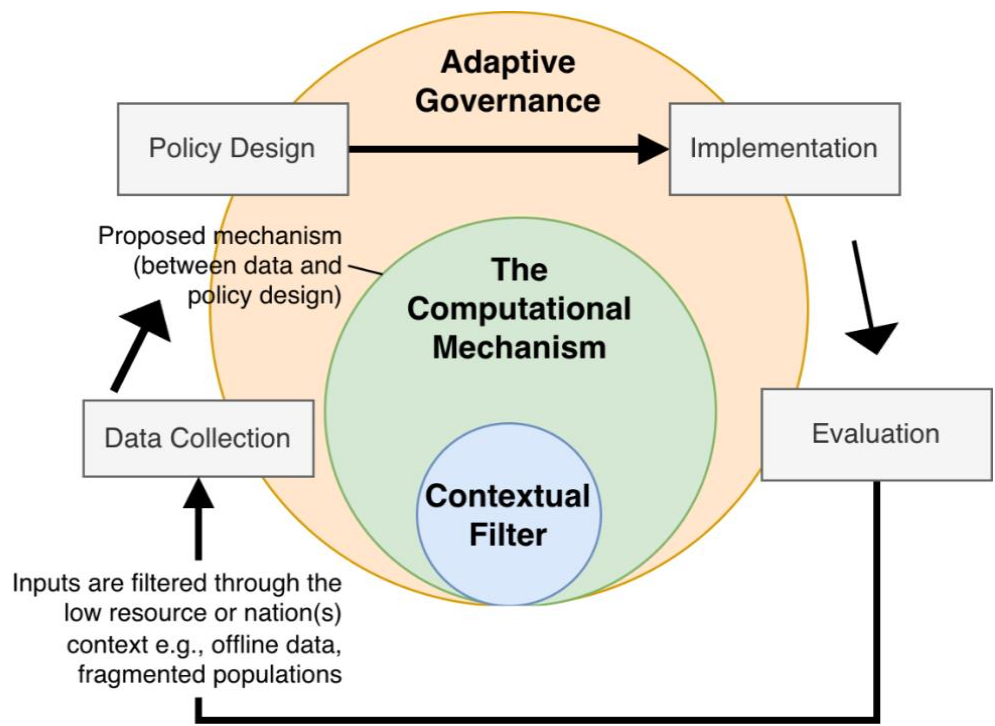

Fig. 1. The Predictive Simulation Conceptual Model for Adaptive Digital Governance
This conceptual model illustrates how the study's predictive simulation workflow (central block) supports adaptive governance theory. By ingesting fragmented data from the targeted nation context and subjecting it to machine learning analysis, the framework generates *ex-ante* policy scenarios. These scenarios function as a feedback mechanism, allowing policymakers to "test" interventions virtually before deployment, thereby bridging the gap between theoretical policy goals and the realities of implementation in resource-constrained environments.

## 3. Methodology

A multi-phase, quantitative research design was employed to support policy experimentation through predictive modeling and simulation. This approach aligns with emerging best practices in digital governance research, which advocate for the integration of descriptive, predictive, and evaluative analytics to inform inclusive public service delivery (Caravaggio et al., 2025; Schmitz, 2025). Designed for application in resource-constrained settings, such as rural and remote areas of the Pacific Islands, the methodology sequentially applies statistical profiling, machine learning techniques, and scenario-based simulation to generate actionable insights for DFL interventions, as presented in Fig. 2.

This framework utilizes supervised machine learning models to identify key policy levers impacting DFL across various demographic, socio-economic, and behavioral dimensions. In the final phase, these levers informed scenario-based simulations that forecasted the effects of targeted interventions, allowing for projections of DFL improvements. This approach not only evaluated which interventions are most effective but also clarified for whom they are most beneficial and under what contextual conditions they succeeded. As a result, the study offers a scalable, data-driven framework for designing, testing, and optimizing DFL strategies aimed at bridging the digital financial divide.

*3.1. Data Source and Preparation*

To facilitate generalizability to other low-resource governance contexts, this study utilizes a high-granularity dataset that serves as a representative archetype for digital behavior in data-sparse, archipelagic, and underserved environments. The analysis draws on secondary data from the UNCDF Pacific Digital Economy Program, encompassing 10,108 respondents across seven Pacific Island nations: Fiji, PNG, Samoa, Solomon Islands, Timor-Leste, Tonga, and Vanuatu (Table 1) (United Nations Development Programme, 2025). While the primary data collection occurred in 2022, this dataset retains high contemporary relevance as the designated empirical baseline for the 2025 UNCDF Pacific Digital Economy Programme (PDEP). It represents the foundational benchmark against which regional policy progress is actively measured, ensuring the analysis supports the current legislative and developmental cycle rather than merely offering a retrospective view.

The original survey employed a stratified random sampling design to facilitate coverage across diverse urban, rural, and maritime zones. However, post-stratification weights were not applied in the source dataset, as the primary objective of the initial collection was exploratory statistical profiling rather than census-adjusted population enumeration. Consequently, this study utilizes unweighted estimators, prioritizing the identification of robust multivariate behavioral relationships over absolute population inference.

As outlined in Table 2, the data captures multidimensional indicators that inform the composite DFL score. The DFL construct is presented as a cumulative index ranging from 0 to 52, formed through the additive aggregation of three distinct competency domains (Equation 1). The score is translated to 0 to 100% in this study for ease of interpretation. This approach is justified by the theoretical consensus that DFL is an integrated capability; functional proficiency requires the simultaneous possession of technical skills, traditional financial knowledge, and digital-financial behavioral readiness (Mikalef et al., 2022; Zheng & Shuang, 2025). The instrument was validated for internal consistency and cultural relevance to capture complex digital behaviors through granular, domain-specific variables.

$$DFL = DC + FC + DFC \quad \text{(equation 1)}$$

DFL = Digital financial literacy
DC = Digital Competency (e.g., device ownership, app usage frequency)
FC = Traditional Financial Competency (e.g., numeracy, budgeting habits)
DFC = Digital Financial Competency (e.g., mobile money usage, scam awareness)

Table 1. Number of participants and baseline DFL score in the dataset

| Country | Number of Participants | Average DFL Score in % |
|---|---|---|
| **Fiji** | 1,678 | 50.9% |
| **PNG** | 1,587 | 41.0% |
| **Samoa** | 1,216 | 43.5% |
| **Solomon Islands** | 1,540 | 41.9% |
| **Timor-Leste** | 1,631 | 39.6% |
| **Tonga** | 1,227 | 44.2% |
| **Vanuatu** | 1,229 | 44.2% |

Table 2. Structure of the survey dataset for DFL profiling

| DFL Domain | Sub-Domain | Example Variables Collected |
|---|---|---|
| **Demographics** | Gender, Age, Language, Geography | Gender, age group, interview language, area type |
| **Socio-Economic Profile** | Education, Occupation, Income, Memory, Numeracy | Highest education level, main job activity, personal income, comfort with mental calculation, |
| **Digital Profile (D)** | Access, Skills, Usage Frequency | Device ownership, internet access, app usage frequency |
| **Financial Profile (F)** | Financial Behavior and Planning | Budgeting habits, receipt keeping, expense planning |

| **Digital Financial (DF)** | Digital Transactional Behavior, Perceived Impact and Risk, financial anxiety | Use of mobile money, online transfers, digital spending tracking, remembering PINs/passwords, awareness of scams, confidence using DFS, experience with fraud |
|---|---|---|

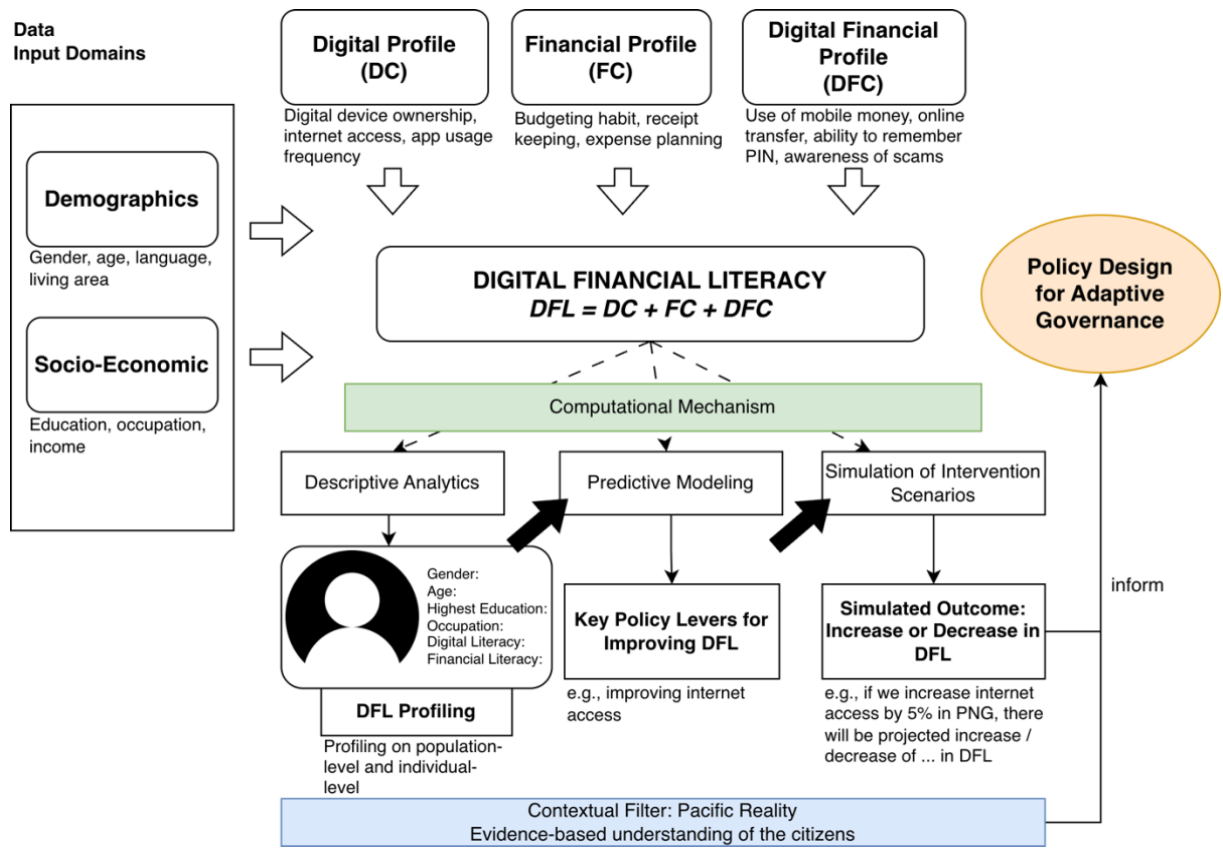

Fig. 2. Implementation of the conceptual model in the case of the Pacific nations data

### 3.2. Phase 1: Descriptive Analytics

The first phase of analysis profiles the structural gaps driving the digital financial divide. While the composite DFL score aggregates three domains: digital competency (DC), traditional financial competency (FC), and digital financial competency (DFC), this analysis prioritizes DFC as the primary lens for profiling. Unlike the digital profile (which measures access and potential) or the financial profile (which measures theoretical knowledge), DFC represents the operational intersection where digital skills are applied to financial tasks. It serves as the proxy for "active engagement." A high digital profile alone does not demonstrate inclusion if the user cannot navigate a banking interface; similarly, high financial knowledge is insufficient without the digital skills to apply it.

From a statistical perspective, the prioritization of DFC is justified by its superior discriminant power compared to the composite DFL score. An analysis of the Coefficient of Variation (CV) reveals that DFC consistently exhibits higher relative variability than total DFL (e.g., in PNG, the DFC variability is approximately 46% compared to 35% for DFL). Details of this analysis are provided in Appendix A. This indicates that while basic digital access and traditional financial knowledge are becoming commoditized, competency remains the primary source of inequality. Therefore, DFC is the strongest correlate of overall DFL and the most critical indicator of a citizen's readiness to engage with digital government platforms.

The analysis involved a distributional and comparative analysis of DFC across gender, age, location, education, and income groups. Frequency distributions, bar plots, and descriptive statistics were used to detect intra-country disparities and cluster characteristics of low-DFL populations. As demonstrated in related studies, such segmentation is critical for informing personalization in digital inclusion strategies (Prasad et al., 2018; Schmitz, 2025). This phase established the groundwork for model training by identifying relevant feature sets and prioritizing subgroups (e.g., rural women with limited digital access) for further analysis.

### 3.3. Phase 2: Predictive Modeling

To identify the key policy levers of DFL, supervised machine learning models were trained and evaluated: Linear Regression, Random Forest, and Gradient Boosting. These architectures were selected for their prevalence in comparative policy analytics and their varying trade-offs between complexity and interpretability (Romanov et al., 2025; Sprenkamp et al., 2025; Young et al., 2022).

#### 3.3.1. Model Selection Criteria: Stability and Transparency

The selection of the base simulation model was governed by a rigorous validation protocol designed to ensure both statistical generalizability and administrative accountability.

To ensure analytical rigor and prevent data leakage, a preprocessing protocol was implemented within a nested validation framework. The dataset was first partitioned into an 80% training set and a 20% hold-out test set, utilizing a stratified split by country to ensure that smaller Pacific nations were adequately represented in the evaluation phase. Within the training phase, a 10-fold cross-validation strategy was employed. Crucially, to mitigate look-ahead bias, all data transformations were encapsulated within the cross-validation pipeline. Specifically, multiple imputation parameters for missing values and one-hot encoding maps for categorical variables were fitted exclusively on the training folds of each iteration before being applied to the validation fold. This isolation allows the model's performance metrics to reflect genuine generalization capability rather than artifacts of the test data (Liang et al., 2025; Nasseef et al., 2022).

The final model selection was guided by three critical metrics relevant to public-sector deployment. First, predictive accuracy ($R^2$) was assessed to determine the model's capacity to explain variance in DFL scores. Second, model stability ($R^2$ variability), measured as the standard deviation of cross-validation $R^2$, served as a proxy for algorithmic equity. Low variability indicates a reliable model that performs consistently well across all population subgroups (e.g., rural vs. urban), whereas high variability signals volatility and exclusion risk, where a model might fail for marginalized groups despite high aggregate accuracy. Finally, algorithmic transparency was prioritized in alignment with Caravaggio et al. (2025), ensuring the selected model offers explainable decision logic, a non-negotiable requirement for public sector accountability.

#### 3.3.2. Feature Importance Analysis for Identification of Policy Levers

Following model selection, a feature importance analysis was conducted to identify the policy levers. The model's features were classified into two distinct categories:

- Modifiable Predictors (Policy Levers): These are variables that government intervention can directly influence. They represent the individual's digital (D), traditional financial (F), and digital financial (DF) characteristics, e.g., digital device ownership, budgeting habits, and digital spending tracking. Crucially, the relationship between these levers and DFL scores is structural. Since the DFL index is constructed as a composite of DC, FC, and DFC capabilities, these modifiable behaviors act as constituent components of the score itself. Therefore, an intervention that successfully alters a behavioral input, such as enabling a user to track spending digitally, is structurally associated with an increase in the DFL score.
- Non-Modifiable Predictors (Segmentation Variables): These include demographic and socio-economic factors such as age, gender, and geographic location. While policy cannot alter these variables (e.g., a government cannot change a citizen's age), they are essential for adaptive service targeting. Their relevance lies in guiding how an intervention is delivered, for example, tailoring outreach methods specifically for rural women, rather than being the target of the change itself.

These distinctions have direct implications for government information policy. Interventions designed to amplify high-impact, modifiable behaviors, such as promoting digital expense-tracking apps or providing access to entry-level devices, can be more effective and scalable. Meanwhile, non-modifiable variables can guide adaptive service targeting without diverting resources toward inelastic attributes.

This classification framework allows the subsequent simulations (Phase 3) to focus on actionable levers while respecting the population's structural constraints.

*3.4. Phase 3: Simulation-Based Policy Scenario Analysis*

The final phase simulated intervention scenarios to forecast the population-level changes of targeted literacy initiatives. Static scenario modelling was implemented using the best-performing predictive model from Phase 2. Synthetic modifications were introduced to the policy levers (e.g., converting digital inaccessibility into full access) to simulate changes in DFL scores. This scenario-based forecasting draws on best practices from related works, which operationalise scenario-building to evaluate governance innovations under constrained conditions (Jafarian et al., 2025; Zheng & Shuang, 2025).

It is important to define the methodological boundaries of these simulations. While this study utilizes cross-sectional data, limiting definitive causal inference, the simulation assumes structural stability in the identified associations to provide a necessary heuristic for prioritization in the absence of longitudinal data. This approach relies on the premise that the strong structural links observed (e.g., between device ownership and literacy) serve as structural stability in the identified associations to provide a necessary heuristic for prioritization in the absence of longitudinal data. Consequently, while Randomized Controlled Trials (RCTs) would be required to rule out selection bias definitively, they are often resource-prohibitive in developing contexts. Therefore, this framework functions as a rigorous, *ex-ante* prioritization tool to guide decision-making before public resources are committed.

Crucially, to facilitate algorithmic fairness, comparative assessments extended beyond aggregate population means to include subgroup performance analysis. The simulation outcomes were disaggregated by key vulnerability dimensions, gender, geography, and socio-economic status, to screen for disparate impact. This step verified that the intervention scenarios did not inadvertently widen the gap for marginalized groups or exhibit unequal error rates between rural and urban populations. This approach allowed for the estimation of marginal gains per intervention while assessing equitable predictive validity, thereby aiding resource allocation that prioritizes equity alongside efficiency.

**4. Empirical Analysis and Findings**

This section presents the findings of the empirical investigation, tracing the study's methodological path from descriptive profiling to predictive simulation. It starts by exploring the digital financial divide through the structural gaps in DFC across diverse populations. These identified gaps inform the subsequent predictive modeling, which establishes the baseline machine learning model and its critical policy levers. Finally, this section translates these predictive insights into a range of simulated intervention scenarios and analyzes the resulting simulations.

*4.1. Descriptive Profiling of Digital Financial Divide*

To understand the mechanics of exclusion, the first phase of analysis delineates the structural gaps underlying the digital financial divide. As established in Section 3.2, DFC is utilized as the strongest correlate of overall literacy and the definitive proxy for engagement with digital government ecosystems. This subsection disaggregates the competency divide across four key stratifications: national population baselines, demographic identifiers, socio-economic constraints, and cognitive-behavioral patterns.

*4.1.1. Population-Level Baseline Insights*

At the population level, a comparative analysis across the seven Pacific Island nations highlights significant structural disparities. To establish a policy-relevant baseline, statistical analysis was done for both the sub-component DFC and the total DFL score (Appendix A). Figure 3 illustrates the strong positive correlation between DFC and total DFL scores across national means. This confirms that DFC acts as the primary driver of overall literacy; nations with higher competency scores consistently achieve higher overall inclusion metrics.

Marked disparities are evident across the region. Tonga and Fiji emerge as regional leaders in digital financial readiness. According to the data, Tonga records the highest mean DFC (46.2%), followed by Fiji (43.7%). These elevated scores suggest that residents in these nations possess not only access but also the foundational capability to leverage DFS tools effectively, factors that align with broader national infrastructure and policy support. In contrast, PNG presents the lowest mean DFC (32.6%) and one of the lowest mean DFL scores (41.1%). This analysis points to a dual-layered exclusion, limited digital access combined with low applied competency, that constrains PNG's population from participating meaningfully in the digital financial ecosystem. The findings underscore that, for lower-performing nations, interventions must address both the "competency gap" and the "access gap."

Median scores across most countries closely mirror their means, particularly in Tonga, Fiji, and Samoa, indicating a relatively balanced distribution of competency. However, analysis of standard deviations reveals substantial intra-country variation, particularly in Vanuatu (17.6%) and Fiji (16.4%). These deviations suggest the coexistence of highly proficient elites alongside severely underprepared populations within the same national context.

The observed ranges of individual DFC scores emphasize this extreme heterogeneity. The data spans from an absolute minimum of 0.0% (observed in Samoa) to a maximum of 92.3% (observed in Fiji). Statistically, this wide dispersion indicates that national averages mask deep inequalities; while Samoa's average DFL is robust (43.3%), the presence of individuals with 0% DFC reveals pockets of total exclusion. For policy, this range is significant because it represents a functional threshold for digital citizenship. An individual with a 0% DFC score is not merely "less skilled"; they are effectively invisible to the digital state, unable to access e-government platforms or digital welfare without intermediation. Moving citizens across this threshold, from zero to basic functional competency, yields a higher return for governance inclusion than incrementally improving the skills of the already proficient. This validates the need for adaptive, segmented targeting, as a one-size-fits-all strategy would fail to address the distinct needs of the completely excluded versus the digitally emerging.

Collectively, these findings justify the need for predictive analytics. By identifying that the divide is rooted in competency (DFC) and characterized by deep internal inequalities, the study validates the development of simulation tools to model how specific interventions can lift these competency scores for the most excluded segments.

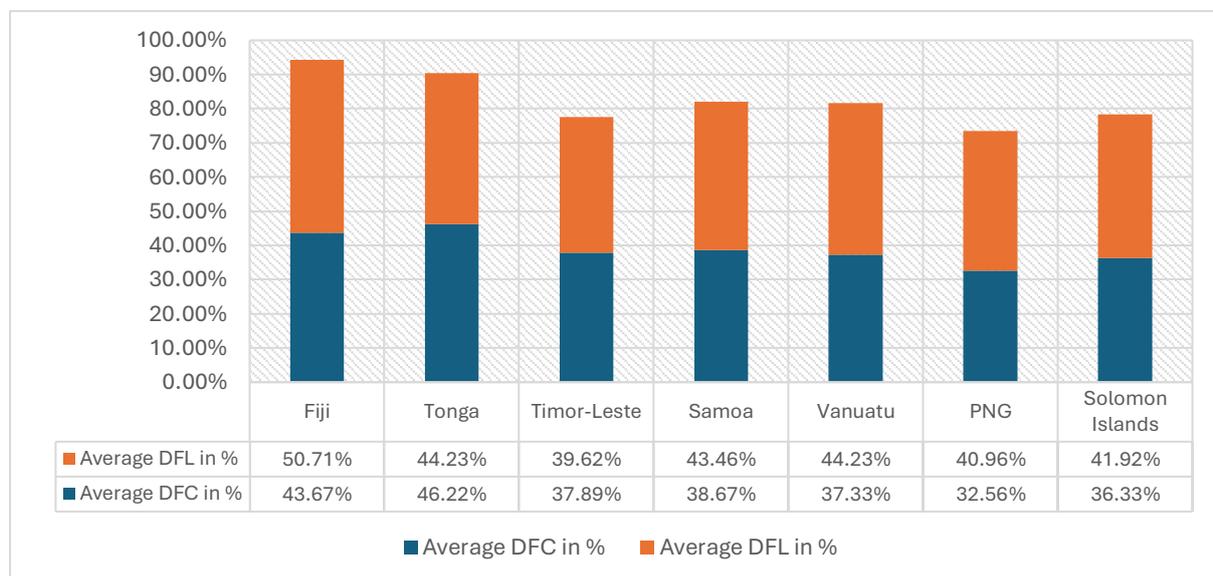

Fig. 3. DFL and DFC Score Distribution by Country

*4.1.2. Demographic Segmentation of Digital Financial Divide*

The findings offer a perspective on how various demographic factors contribute to disparities in DFC across Pacific Island communities. Table 3 outlines disparities in DFC across demographic groups. The divide is characterised based on spoken language, residence area, household composition, age group, their local access to DFS, and gender.

The descriptive analysis reveals that language and geography function as proxies for deeper structural and economic realities within the Pacific region. The most pronounced disparity is linguistic: Tongan speakers achieve a DFC score of 46.4%, whereas Tok Pisin speakers (primarily in PNG) average only 34.1%, a substantial 12.3% gap.

This divergence reflects distinct national contexts. Tonga is a compact Polynesian nation characterized by a highly integrated remittance economy, where a significant portion of the population relies on financial inflows from the diaspora. This economic structure necessitates the routine use of digital transfer platforms, effectively mandating a baseline of digital literacy for daily survival. This interpretation is supported by the socio-economic data (Table 4), which shows that regular overseas workers, the primary drivers of remittances, achieve one of the highest competency scores (44.7%). In contrast, Tok Pisin is the lingua franca of PNG, a nation defined by rugged, archipelagic fragmentation and a larger informal cash economy. Here, the language gap captures the friction of accessing digital services in a fragmented market where physical cash remains dominant due to infrastructural barriers.

Similarly, the urban-rural divide (42.4% vs. 38.0%) highlights the impact of service density. Urban residents benefit from the agglomeration effect, proximity to bank branches, consistent electricity, and 4G networks, which facilitate the routine practice required to maintain digital skills. This gap widens in hyper-rural contexts; individuals in hamlets with populations under 100 score just 36.6% compared to 41.4% in larger towns. This gradient confirms that digital competency is environmentally conditioned; without the physical infrastructure to practice financial transactions, skills decay.

Finally, social structure plays a functional role in skill acquisition. Individuals living alone record significantly lower DFC scores (38.5%) compared to those residing with family members (>40%). This suggests that communal financial management acts as an informal learning mechanism, where digital skills are transferred intergenerationally (e.g., younger members assisting elders). Interestingly, the absence of a significant gender gap (Male 39.9% vs. Female 39.8%) contrasts with global trends. In the Pacific context, women often function as the primary managers of household finances, a role that necessitates a functional level of digital proficiency that counterbalances other structural exclusions.

Table 3. Disparities in digital financial competency across demographic groups

| Category | Lowest DFC | | Highest DFC | | Gap % |
| --- | --- | --- | --- | --- | --- |
| | Value | DFC % | Value | DFC % | |
| **Spoken Language** | Tok Pisin | 34.10% | Tongan | 46.40% | 12.30% |
| **Living Area** | Living in a village, hamlet or other community with fewer than 100 people | 36.60% | Living in a town or city with between 3,000 and 100,000 people | 41.40% | 4.80% |
| **Residence Area** | Rural | 38.00% | Urban | 42.40% | 4.30% |
| **Household Composition** | Entirely alone | 38.50% | With the family that brought you up | 40.60% | 2.10% |
| **Age Group** | Age: 65-74 | 38.50% | Age: 25-34 and Age: 35-44 | 40.40% | 2.00% |
| **Local Access to Services** | A neighbor | 39.40% | ATM | 40.90% | 1.40% |
| **Gender** | Female | 39.80% | Male | 39.90% | 0.10% |

*4.1.3. Socio-Economic Segmentation of Digital Financial Divide*

The findings underscore the role of socio-economic factors in shaping DFC levels, with occupation, employment sector, education, income, and certain personal characteristics contributing to notable disparities, as outlined in Table 4.

Formal education emerges as the strongest structural predictor of DFC, creating a stark 13.3% capability gap between those with no formal schooling (31.7%) and those with postgraduate qualifications (45.0%). This gradient suggests that DFC is heavily dependent on foundational cognitive skills: literacy, numeracy, and abstract reasoning, which are cultivated in formal learning environments. Notably, the data reveal a threshold effect at the high school level; graduates achieve a score of 40.8%, suggesting that secondary education provides the minimum cognitive scaffolding necessary to navigate complex digital financial interfaces, even without specialized tertiary training.

Occupational patterns further illuminate the influence of the Pacific's remittance-dependent economy on digital skills. Regular overseas workers, individuals integrated into labor mobility schemes, demonstrate high competency (44.7%), significantly outperforming informal or subsistence workers categorized as "Other" (35.9%). This disparity is structural rather than merely occupational; overseas workers are often compelled to use digital platforms for cross-border remittances, creating a forced adoption effect that builds competency through necessity. In contrast, the agricultural and fishing sectors, which remain cash-dominant and localized, record the lowest scores (38.2%). This implies that without the environmental pressure to transact digitally, whether for work or remittances, digital skills atrophy or fail to develop.

Finally, income and mobility aspirations act as material and motivational enablers. A clear income gradient exists (6.5%), where earning a moderate income (USD 385–480/fortnight) correlates with higher DFC (42.6%) compared to those with no income (36.1%). This likely reflects the ability to afford the recurring costs of digital participation (data plans and device maintenance) required for experiential learning. Interestingly, individuals expressing an intent to migrate permanently score significantly higher (42.7%) than those without such plans. This suggests an anticipatory adaptation effect, where individuals proactively upskill digitally to prepare for integration into modernized economies abroad. For policymakers, this highlights that digital exclusion is often a function of economic isolation; interventions must therefore mimic the practice effects of formal employment and migration to be effective for the stationary, informal workforce.

Table 4. Disparities in digital financial competency across socio-economic groups

| Category | Lowest DFC | | Highest DFC | | Gap |
|---|---|---|---|---|---|
| | Value | DFC % | Value | DFC % | |
| **Highest level of education** | No formal education | 31.70% | Postgraduate | 45.00% | 13.30% |
| **Occupation** | Other | 35.90% | A regular overseas worker | 44.70% | 8.80% |
| **Employment Sector** | Agriculture, farming, or fishing or forestry | 38.20% | Technology | 45.30% | 7.00% |
| **Income** | No income | 36.10% | Between $385 and $480 per fortnight | 42.60% | 6.50% |
| **General characteristics** | I have a health condition or disability that limits my day-to-day activities | 38.30% | I intend to move overseas permanently | 42.70% | 4.40% |

*4.1.4. Behavioral and Cognitive Segmentation of Digital Financial Divide*

While the preceding demographic and socio-economic analyses identified who is excluded, this section examines how exclusion operationally manifests. Unlike structural variables (e.g., age or income), which serve as external correlates to competency, behavioral and cognitive factors act as constitutive components of the DFC score itself. Analyzing these dimensions is critical for government information systems because it transitions the focus from static profiling to functional diagnosis. By unraveling the specific user actions, such as security protocols, payment habits, and risk perception, that directly determine the DFC score, this analysis isolates the actionable "mechanisms of use" that policy can target through system design and education, rather than just resource allocation. Table 5 reveals notable disparities in DFC among various groups, illustrating how beliefs, ownership, behaviors, safety practices, and payment preferences impact the digital financial gap.

Behavioral patterns, specifically regarding digital hygiene and risk management, emerge as the most profound differentiators of competency, surpassing even demographic factors. The single largest gap in the dataset (17.2%) exists between users who actively verify website security (56.6% DFC) and those who rely on analog security methods like writing down passwords (39.3%). This suggests that high competency is not merely about execution but involves a sophisticated mental model of digital risk. Similarly, the ability to identify and delete phishing emails correlates with a 13.4% advantage over those who engage with fraudulent links. This indicates that "digital street smarts", the ability to discern threat from opportunity, is a critical functional threshold for safe inclusion.

Transactional habits reveal a distinct experiential learning mechanism. A stark 16.7% gap separates card users (44.4%) from those with no payment method (27.8%), identifying a cash trap where lack of access inhibits skill development. This confirms that digital competency is reinforced through a virtuous cycle of use: the daily friction of navigating card terminals, managing PINs, and tracking digital balances acts as a continuous micro-training regimen. Without this recurring interaction, cash-dependent users are deprived of the feedback loops necessary to build confidence and fluency, causing their skills to stagnate at the baseline.

Finally, cognitive frames act as gatekeepers to inclusion. A significant 11.1% gap exists between individuals who perceive DFS as a tool for economic efficiency (keeping costs down) and those who view it through an exclusionary lens (designed for men). This demonstrates that perception precedes competence; users who fail to see the utility of digital finance, or who feel culturally alienated by it, self-select out of the ecosystem before skills can develop. Furthermore, barriers such as lack of identification or distrust in institutions correlate with lower scores, highlighting that low DFC is often a symptom of broader administrative exclusion, where citizens are locked out of the formal identity systems required to even attempt digital participation.

Table 5. Disparities in digital financial competency across digital financial groups

| Category | Lowest DFC | | Highest DFC | | Gap |
| --- | --- | --- | --- | --- | --- |
| | Value | DFC % | Value | DFC % | |
| **DFS behaviours** | Keep a record of pin numbers or passwords for financial services | 39.30% | Check that a website is secure before entering payment details | 56.60% | 17.20% |
| **Payment method beside cash** | Nothing | 27.80% | Used card | 44.40% | 16.70% |
| **Safety Response on Potential Email Fraud** | Follow the link | 36.10% | Delete the email | 49.50% | 13.40% |
| **DFS beliefs** | Are designed for men more than women | 31.70% | Keep costs down for small businesses | 42.80% | 11.10% |
| **DFS ownership** | Cryptocurrency | 40.80% | A digital wallet | 44.10% | 3.30% |
| **Reason for only using cash** | Another reason | 33.40% | Prefer to use cash | 36.60% | 3.20% |
| **Primary method of managing money** | No money usage, not even cash | 33.30% | Cash only | 36.10% | 2.80% |

## 4.2. Predictive Modeling of Key Policy Levers

Following the identification of the structural and behavioral contours of the digital financial divide, the second phase of analysis moves from diagnosis to predictive analysis. This section identifies the optimal machine learning architecture to serve as the study's intervention baseline. The objective is twofold: to select a model that satisfies the dual governance requirements of predictive accuracy and algorithmic transparency, and to extract the specific, modifiable policy levers that will drive the DFL improvements simulated in the final phase.

### 4.2.1. Model Performance

Upon applying the validation protocol outlined in Section 3.3.1, the predictive performance of the three candidate architectures was evaluated. As summarized in Table 6, Linear Regression emerged as the most robust model for policy simulation.

Linear Regression achieved the highest predictive accuracy on the test set ($R^2$ = 95.89%), outperforming both Gradient Boosting (94.30%) and Random Forest (86.88%). Furthermore, it demonstrated superior error minimization, recording the lowest Root Mean Squared Error (RMSE: 1.43) and Mean Absolute Error (MAE: 1.10). While the other models demonstrated moderate-to-high explanatory power, they exhibited higher error margins, indicating a lower fidelity in capturing the linear drivers of competency. Crucially, Linear Regression exhibited the highest stability, with a minimal cross-validation standard deviation (CV $R^2$ Std: 0.0008). This is significantly lower than the volatility observed in Gradient Boosting (0.0029) or Random Forest (0.0069). Statistically, this low deviation translates to a tight 95% confidence interval for the model's predictive power (95.8%±0.16%). This effectively acts as a sensitivity analysis, confirming that the Linear Regression model is robust to data perturbations and performs consistently across the diverse demographic segments of the Pacific population.

Consequently, Linear Regression was selected as the baseline model for the simulation phase. This choice is justified not only by its superior stability and accuracy but also by its operational transparency. Unlike black box ensembles, the linear model provides transparent coefficients (Equation 2) that allow policymakers to explicitly quantify the marginal effect of specific interventions, such as providing a smartphone, while mathematically holding demographic factors constant. For policymaking purposes, the model must be able to assist policymakers in justifying decisions based on clear, traceable evidence. The model's interpretability enhances its communicability, allowing policymakers to clearly trace how specific demographic, digital, or financial behaviors influence literacy outcomes. Its application in this study provides a replicable and scalable workflow for digital financial inclusion forecasting, offering a valuable decision-support tool for adaptive policy planning in underserved and digitally vulnerable contexts.

Table 6. Comparative performance of machine learning models for predicting DFL

| Machine Learning Model | MSE | RMSE | MAE | Test $R^2$ | Cross-validation $R^2$ mean | Cross-validation $R^2$ std |
|---|---|---|---|---|---|---|
| **Linear Regression** | 2.04 | 1.43 | 1.1 | 95.9% | 95.8% | 0.0008 |
| **Random Forest** | 6.53 | 2.55 | 1.97 | 86.9% | 85.9% | 0.0069 |
| **Gradient Boosting** | 2.84 | 1.68 | 1.29 | 94.3% | 94.1% | 0.0029 |

Note: $R^2$ = Coefficient of Determination; MAE = Mean Absolute Error; RMSE = Root Mean Squared Error; MSE = Mean Squared Error; CV = Cross-validation.

$$DFL = DFL_{Base} + (weight \; x \; factor) \quad \text{(equation 2)}$$

*4.2.2. Key Predictors: Policy Levers*

The feature importance analysis derived from the best model provides insights into the behavioral and capability domains most strongly associated with DFL (Table 7). Among digital attributes, device ownership, such as tablets, laptops, and desktop computers, emerged as the most influential policy-relevant predictor (5.46%). This underscores the enduring role of device access as a prerequisite for digital engagement and suggests that closing hardware access gaps remains foundational to inclusive digital financial ecosystems.

Closely following were the digital content creation (4.41%) and computational skills. These findings highlight that basic productivity skills, not merely access, are pivotal to enabling functional use of digital financial tools, reinforcing the need for interventions that go beyond device provision to include practical digital skill development.

In the domain of financial attributes, routine expense recording, including keeping receipts or recording expenses (1.40%) and planning income or expenditures in advance (1.11%), was found to moderately influence DFL. These are direct targets for financial literacy programs and may be embedded in school curricula, community workshops, or fintech applications aimed at reinforcing sound financial habits. By contrast, attitudinal indicators, such as optimism regarding financial security over the next five years (0.72%), were shown to have limited predictive value, suggesting that interventions focused on behavioral reinforcement are likely to yield more tangible outcomes than those aimed solely at improving financial confidence.

Crucially, digital financial attributes had the most pronounced influence. The ability to digitally track personal spending emerged as the single most impactful predictor across all categories (13.70%). This capability not only reflects a convergence of digital and financial skills but also signals readiness for autonomous financial management within digital ecosystems. Similarly, independent digital management of personal finances (4.79%) significantly enhanced DFL, indicating that competencies enabling users to take control of their digital financial environment should be central to intervention design. Conversely, while the experience of avoiding online scams or phishing attempts (0.70%) had relevance, it played a comparatively minor role, possibly reflecting a reactive rather than enabling aspect of digital engagement.

These results suggest that policy levers aimed at expanding access to smartphones and laptops, embedding digital budgeting tools into DFS interfaces, and promoting foundational digital skill-building are likely to have the most substantial returns in lifting DFL outcomes. As such, public programs that integrate digital and financial capability development, particularly through community centers, mobile learning apps, or agent-based delivery models, may represent the most impactful pathway to bridging the digital financial divide in Pacific Island nations. This finding informs resource allocation: governments should prioritize hard infrastructure (devices) and functional tools (budgeting apps) over purely educational campaigns aimed at shifting attitudes.

Table 7. Modifiable Behavioral Drivers and Policy Importance Scores

| Domain | Targeted Policy Lever (Behavior/Asset) | Relative Predictive Weight |
|---|---|---|
| **Digital** | | |
| Infrastructure | Device Ownership (Tablet, Laptop, or Desktop) | 5.46% |
| Skills | Digital Content Creation (Editing Docs/Images) | 4.41% |
| Skills | Computational Skills (Spreadsheet Formulas) | 3.26% |
| **Financial** | | |
| Behavior | Expense Recording (Keeping Receipts) | 1.40% |
| Planning | Budget Management (Planning Income/Expenses) | 1.11% |
| Attitude | Financial Optimism (5-Year Security Outlook) | 0.72% |
| **Digital Financial** | | |
| Behavior | Digital Spending Tracking (App/Online Monitoring) | 13.70% |
| Behavior | Digital Autonomy (Independent Management) | 4.79% |
| Safety | Cybersecurity Resilience (Scam Avoidance) | 0.70% |

Note: Relative Predictive Weight represents the percentage contribution of each feature to the model's total predictive power, derived from the standardized coefficients of the Linear Regression model. Targeted Policy Levers correspond to specific DFL behavior items: Digital Spending Tracking (using digital tools/apps) is distinct from manual Expense Recording (keeping physical receipts); Device Ownership refers specifically to computers or tablets (excluding basic mobile phones); Computational Skills denotes the use of spreadsheet formulas; and Digital Autonomy refers to the ability to execute transactions without third-party assistance.

*4.3. Simulation of Intervention Scenarios*

The final phase integrated the policy levers into simulated policy scenarios. It is important to note that the results presented below represent forecasted potential gains based on the structural associations identified in the model. They serve as an *ex-ante* heuristic for prioritization, identifying which levers have the highest theoretical probability of success, rather than reporting realized effects from field trials.

*4.3.1. Scenario Outcomes and Strategic Prioritization*

Table 8 details the simulated outcomes for individual and bundled interventions. The scenarios were designed to test specific policy levers (e.g., "Device Access") when delivered as strategic bundle programs or individual intervention programs. The analysis reveals a clear hierarchy of effectiveness, distinguishing between foundational drivers and incremental nudges.

1. The "Digital-First" Bundle: The simulation identifies the "Digital Capability Bundle", which combines device ownership, digital content creation, and computational skills, as the single most effective strategy. This scenario is forecasted to yield a 5.5% increase in DFL scores, positively impacting 57% of the population. Interestingly, this targeted bundle outperformed the more complex "All-9 Capabilities" bundle (5.3%). This suggests a principle of diminishing returns: creating a focused digital foundation is more efficient than diluting resources across every possible financial and digital competency simultaneously.

2. Infrastructure as the Prerequisite: Supporting the "second-level digital divide" theory, the simulation confirms that device ownership alone is the most powerful standalone lever. Simply providing access to smartphones or computers is projected to lift DFL by 4.7% across 60% of the population. This underscores that physical infrastructure is the non-negotiable entry ticket; without the device, behavioral interventions have no platform on which to operate.

3. Behavioral Interventions as Scaffolding: In contrast, interventions focused solely on daily behaviors, such as digital spending tracking (2.9%) or cybersecurity resilience (3.1%), show more modest projected gains when delivered in isolation. This indicates that behavioral nudges lack traction without the preceding exposure to digital tools. For government information systems, this implies a sequencing requirement: behavioral modules should serve as scaffolding built on top of device access programs, rather than as standalone solutions.

4. Feasibility and Cost-Effectiveness Implications: While the "Digital Capability Bundle" offers the highest theoretical return (5.5%), practitioners must weigh this against operational feasibility and fiscal constraints. A program providing hardware to 60% of the population carries a significantly higher capital cost than a "Safety Awareness" campaign (3.1% gain). However, the simulation reveals that the low-cost option yields a correspondingly low impact. This finding supports a tiered investment strategy: high-cost infrastructure interventions should be reserved for the high-leverage segments identified in Section 4.3.2 (e.g., caregivers), while lower-cost behavioral nudges can be scaled broadly once the infrastructure baseline is met.

Table 8. Projected population-level outcomes of simulated interventions

| Intervention Scenario | Reached Population in % | Forecasted DFL increase |
|---|---|---|
| **Strategic Bundles (Combined Levers)** | | |
| Comprehensive Bundle (All 9 Capabilities) | 40.7% | 5.3% |
| Digital Capability Bundle (Device + Content + Computation) | 57.0% | 5.5% |
| Financial Capability Bundle (Budgeting + Recording) | 40.0% | 3.2% |
| Digital Financial Safety Bundle (Autonomy + Cybersecurity) | 38.8% | 3.0% |
| **Individual policy levers** | | |
| *Digital infrastructure and skills* | | |
| 1. Device Access (Smartphone/Computer) | 60.0% | 4.7% |
| 2. Digital Content Creation (Skills) | 49.0% | 3.7% |
| 3. Computational Skills (Spreadsheets) | 49.8% | 3.3% |
| *Financial Behavior* | | |
| 4. Financial Optimism (Attitude) | 52.6% | 3.3% |
| 5. Budget Management (Planning) | 41.8% | 2.9% |
| 6. Expense Recording (Behavior) | 42.5% | 2.9% |
| *Digital Financial Behavior* | | |
| 7. Digital Autonomy (Independent Management) | 47.4% | 3.1% |
| 8. Cybersecurity Resilience (Safety) | 49.1% | 3.1% |
| 9. Digital Spending Tracking (Monitoring) | 42.5% | 3.0% |

Note: Forecasted DFL Increase represents the projected percentage point rise in the population-wide DFL score under the simulated scenario, calculated using static scenario modeling based on the predictive coefficients from Phase 2. This is an *ex-ante* heuristic, not a realized field trial result. Reach indicates the proportion of the population (N=10,108) currently lacking the specific capability who would effectively receive the intervention in the simulation.

*4.3.2 Equity and Precision Targeting: Identifying the High-Leverage Responder Segment*

Crucially, the simulation revealed who is forecasted to benefit most, enabling precision targeting. The analysis identifies distinct population segments that show the highest projected responsiveness to the modifiable policy levers. These findings point to several key recommendations for refining and scaling DFL initiatives for these high responder profiles, as shown in Table 9.

The simulation projects that interventions will yield a positive DFL increase primarily among groups with low baseline scores, specifically the "High-Leverage" Responder segment (Average Forecasted Gain: +3.62%). This broad segment comprises young women (aged 15–34) in informal caregiving roles, mostly from PNG and Timor-Leste. Despite having limited formal income and often identifying as "housewives," these individuals are digitally connected via smartphones and highly motivated to manage their household budgets. Their responsiveness challenges the assumption that economic inactivity equates to digital passivity. Geographically, this group is concentrated in rural areas and small towns (pop. 100–3,000), where baseline infrastructure is emerging but not yet ubiquitous. Consequently, this segment represents a high-ROI target for government policy; tailoring mobile-based micro-learning modules to this demographic could bridge the gender divide more effectively than broad-spectrum programs by leveraging their existing smartphone access and budgeting responsibilities.

Within this broader population, a subset is forecasted to achieve exponential gains, identified as the deep impact outlier (Average Forecasted Gain: +26.3% or 13.7 points). This subgroup shares the demographic traits of the broader responder, young, female, caregiver, but is distinguished by a specific cognitive and motivational profile. Unlike the broader group, these improvers report high comfort with mental calculations and strong memory retention (e.g., for passwords) and reside predominantly in Tonga, Vanuatu, and Samoa, often in semi-urban settings. This phenomenon illustrates the "Caregiver Paradox": while economically excluded from formal labor, these women act as the primary financial managers for their households. The simulation suggests that when their intrinsic motivation (budgeting needs) meets the right intervention (digital tools), their competency scores surge dramatically because they have an immediate use case for the skills.

Table 9 synthesizes the profiles of these two groups to guide intervention design. The alignment between the general responder and the high-improver confirms the validity of targeting young female caregivers. However, the distinction in cognitive profiles suggests that while basic access interventions will help the broad base, achieving the deep intervention impact requires programs that also scaffold cognitive confidence and numeracy. For government information systems, these findings validate a move toward persona-based segmentation. Rather than targeting "the poor" or "rural residents" as monolithic blocks, systems should identify and prioritize the "Connected Caregiver" persona, a digitally accessible but economically invisible demographic that holds the key to household-level digital transformation.

Simulated interventions in the study inform insight for policy design that prioritize women aged 25–34 with informal caregiving roles, particularly in low-income households; Expand focus to include peri-urban and small-town environments where infrastructural gaps coexist with mobile access; Emphasize practical, context-relevant content that builds on learners' motivation, numeracy, and real-life application; Reinforce financial planning and digital confidence as core pillars of training; Engage high improvers as peer trainers or community champions to expand outreach. The alignment between general and high-performing profiles confirms the validity of targeted DFL intervention strategies while highlighting important distinctions.

Table 9. Comparative profile of high-leverage improvers for DFL intervention design

| Category | Broadly Responsive Population Segments | Most Significantly Improved Individuals | Policy Implications |
| --- | --- | --- | --- |
| **Location** | PNG and Timor-Leste | Tonga, Vanuatu, Samoa | Semi-urban areas with emerging infrastructure offer a conducive context for DFL learning. |
| **Language** | Tetum | Tongan, Bislama, Samoan | Local-language delivery supports cultural resonance and accessibility. |
| **Age Group** | 15–34 | 25–34 | Tailored life-stage approaches are key for young adult women. |
| **Gender** | Female | Female | Gender-responsive design should prioritize caregiving and financial management roles. |
| **Highest Education Level** | Upper secondary or high school | Upper secondary or high school | Programs can succeed even with limited formal education. |
| **Employment** | Informal caregiving; no personal income | Informally engaged; limited formal employment | Economic inactivity is not a barrier if interventions target relevant use cases. |
| **Digital Technology Use** | Own smartphones; frequent internet access | Own smartphones; frequent internet access; reside in areas with moderate infrastructure | Digital access is necessary but not sufficient, contextual factors enhance uptake. |
| **Cognitive Profile** | No prominent cognitive traits observed | Motivated, good memory, comfortable with numbers | Intrinsic motivation and cognitive readiness are strong enablers of learning. |
| **Financial Behavior** | Active budgeting; manage household expenses | Active budgeting; manage household expenses; also comfortable with tracking and saving digitally | Behavioral intent matters, DFL growth is linked to proactive financial habits. |
| **Emotional State** | Concerned about finances; anxious about being left behind by tech | Resilient; optimistic learners | DFL programs should address emotional readiness and confidence-building. |

Note: Broadly Responsive Segments are defined as clusters showing a forecasted DFL gain across all scenarios. Most Significantly Improved Individuals ("Deep Impact" outliers) are defined as the top decile of responders, showing highest forecasted gain. Profiles are derived from the subgroup analysis of simulation results disaggregated by demographic variables..

*4.3.3. Adaptive Efficiency: Identifying Non-Responders to Prevent Misallocation*

A critical function of adaptive governance is the capacity to filter out populations where interventions yield diminishing returns. The simulation explicitly identifies a distinct cohort of "non-responders", participants forecasted to show zero or negligible DFL growth under intervention conditions. This group is primarily composed of urban, formally employed professionals, particularly in Fiji and Tonga, who already possess high baseline DFL scores. For example, employed individuals account for the highest volume of projected non-response to the device access intervention, representing approximately 10.1% of the total study population (n=1,022). The data suggests this cohort has hit a "competency ceiling"; because they are already digitally fluent and equipped, basic access interventions (e.g., providing a smartphone) are redundant.

The distribution of non-responders reveals notable geographic saturation. Fiji and Tonga consistently emerge as regions with the highest density of high-baseline users. In Fiji, projected non-response rates range from 7.5% to 12% across varied intervention types, including device access and financial tracking. Similar trends are observed in the Solomon Islands, identifying specific sub-national regions where deploying basic infrastructure interventions would result in policy redundancy. Occupational status serves as the strongest predictor of this ceiling effect, with formal employment correlating most strongly with non-improvement across all six intervention categories.

In a traditional, universal rollout (e.g., "Smartphones for All"), these non-responders would absorb significant resources with no marginal policy gain. By identifying this segment *ex-ante*, the predictive model functions as a fiscal safeguard. It provides the necessary intelligence for government information systems to exclude high-proficiency users from basic subsidy programs, allowing resources to be redirected toward the high-leverage segments identified in Section 4.3.2. This capability transforms the policy approach from a static welfare distribution into a dynamic, efficiency-maximizing investment strategy.

## 5. Discussion

This discussion synthesizes the findings to explore their implications for adaptive governance in the public sector. The results offer empirical support for the transition toward anticipatory, data-enabled policymaking, suggesting that predictive tools can enhance the precision of resource allocation. The analysis is organized into four thematic sections. It begins by interpreting the findings to connect the simulation's predictive utility with broader governance challenges, such as infrastructure prioritization and algorithmic transparency. This is followed by an evaluation of policy implications, specifically regarding the sequencing of digital interventions and the operationalization of precision targeting. Subsequently, the study's theoretical and practical contributions are articulated, framing the computational architecture as a potentially transferable solution for data-constrained environments. Finally, the discussion defines the study's limitations and outlines a trajectory for future research to further evolve this research.

*5.1 Interpretation of Findings*

This study demonstrates how predictive simulation tools can be used to transition digital financial inclusion policy from retrospective assessment to anticipatory, data-enabled governance. By providing empirical validation for the theories of Adaptive Governance, the research establishes a framework that extends beyond its immediate regional application. While the empirical data is grounded in the Pacific Island context, the proposed predictive simulation pipeline, sequencing descriptive profiling, interpretable machine learning, and scenario simulation, offers an analytical framework that is adaptable to diverse geographic settings. Consequently, the framework holds significant transferability for other resource-constrained and archipelagic governance environments seeking to implement evidence-based information systems. The following sections interpret these findings through this lens, highlighting how the study's insights extend, align with, or challenge existing paradigms in digital public administration.

*5.1.1. Operationalizing Feedback Loops in Adaptive Governance*

The primary contribution of this study is the computational model of adaptive governance within resource-constrained environments. While existing theoretical frameworks emphasize the need for institutions to "learn" and "adjust" to local feedback, they often lack the computational mechanisms to do so *ex-ante*. This study bridges that gap by introducing a framework explicitly designed for "low-capacity" settings. Unlike complex digital twin models that require massive real-time data streams, often unavailable in developing nations, this approach demonstrates how periodic national survey data can be repurposed into dynamic intelligence for government information systems.

This low-resource viability enables the creation of critical negative feedback loops required for adaptive systems. For instance, the simulation's projection of "non-improvers" (Section 4.3.3), specifically urban professionals in Fiji and Tonga who reached a competency ceiling, provides a signal to halt investment. In a traditional, static policy model, resources might be allocated universally based on population size, resulting in deadweight loss. However, the predictive model reveals that such an approach is inefficient. These findings align with existing research on the necessity of precision in funding allocation, extending the argument to demonstrate that exclusion (filtering out high-proficiency users) is as vital for system efficiency as inclusion (Caravaggio et al., 2025).. Thus, the predictive tool functions not only as a forecaster but as a fiscal safeguard within the digital bureaucracy, achievable even without real-time data infrastructure.

*5.1.2. The Primacy of Digital Infrastructure over Behavioral Nudges*

The predictive modeling results challenge the prevailing emphasis on purely educational or behavioral interventions in isolation. The feature importance analysis revealed that digital spending tracking (13.7%) and device ownership (5.46%) far outranked attitudinal variables like financial optimism (0.72%). This finding contrasts with behavioral economic theories that prioritize nudges or mindset shifts as primary policy levers. Instead, it aligns strongly with the second-level digital divide literature, which posits that physical access and functional computational skills are prerequisites for higher-level engagement. The simulation results confirm this hierarchy: interventions targeting foundational digital capabilities (a bundle of device access + skills) yielded the highest population-level gain (5.5%), whereas purely financial behavioral interventions yielded significantly lower returns. For government information systems, this dictates a "Digital First" sequence: e-government platforms must first solve the hardware and interface-fluency gap before deploying complex financial literacy content.

*5.1.3. Precision Targeting and the "Caregiver Paradox"*

The study's identification of young women in informal caregiving roles as the highest-leverage segment (High-Improvers) offers robust empirical support for the taxonomy of digital vulnerability (Pérez-Escolar & Canet, 2023). However, the findings refine this theory by revealing a "Caregiver Paradox": this demographic is often economically excluded (low formal income) but digitally connected and cognitively ready. This aligns with observations on the multidimensional nature of literacy but extends them by showing that motivation and life-stage responsibility (e.g., managing household budgets) act as multipliers for intervention success (Zhu, 2025).

For policymakers, this signals that "vulnerability" is not synonymous with "inability." The high responsiveness of this group validates the move toward persona-based policy design, suggesting that government information systems should leverage demographic data to micro-target "mobile-first" micro-learning modules to caregivers, rather than relying on broad-spectrum classroom training, which often fails due to time constraints.

This success in isolating specific behavioral nuances suggests that the predictive simulation logic may hold transferability to other public service domains. The underlying computational workflow, identifying modifiable behavioral drivers (policy levers) to simulate intervention outcomes, is not intrinsically limited to financial literacy. Future research could explore adapting this logic to adjacent fields, such as digital health adoption or e-government service uptake. For instance, similar modeling approaches might assist in identifying behavioral levers for digital identity adoption, positioning the framework as a potentially versatile instrument for evidence-based governance.

*5.1.4. Algorithmic Transparency as a Pillar of Public Trust*

Finally, the model selection process offers critical insights for the ethical deployment of AI in government. The finding that a standard Linear Regression model achieved comparable accuracy to complex "black box" models (like Gradient Boosting), with significantly lower volatility, challenges the assumption that complexity is always superior for policy work (Hemanth Sai et al., 2025; Young et al., 2022; Zhang et al., 2021).

This simplicity supports the transferability of the framework. The model's reliance on standard demographic and behavioral variables—such as device ownership, income source, and digital usage frequency—suggests that this approach could be easily adapted for regions with similar data profiles, such as the Caribbean, Sub-Saharan Africa, or Southeast Asian archipelagos. In these contexts, where data is often fragmented similarly to the Pacific experience, the use of standard, low-complexity inputs reduces the technical barrier for governments wanting to adopt the system.

Furthermore, this approach aligns with the need for transparency in decision-making (de Bruijn et al., 2022; Ripamonti, 2024). In the context of the Pacific Islands, where trust in formal institutions is often fragile, the use of an interpretable model allows policymakers to explicitly explain why a specific demographic is prioritized (e.g., "Device access adds 4.7% to DFL"). This suggests that for government information systems, interpretability is a functional requirement for legitimacy, ensuring that automated recommendations remain explainable, auditable, and transferable across diverse governance landscapes.

*5.2. Policy Implications*

The findings from this study extend beyond the immediate context of the Pacific Islands, offering a scalable blueprint for public sector policy in any digitally underserved or resource-constrained region. The use of predictive analytics in the research supports a fundamental transition in government information systems: moving from retrospective monitoring to anticipatory, evidence-based governance.

*5.2.1. The "Digital-First" Sequencing Protocol*

The simulation results establish a clear hierarchy of intervention effectiveness, challenging current policy sequencing. The finding that foundational digital capabilities (device access + basic skills) are associated with a projected 5.5% DFL improvement compared to negligible returns from purely attitudinal interventions dictates a "Digital-First" sequencing protocol. Governments should avoid deploying complex financial literacy curricula until the "hardware gap" is bridged. Policies must prioritize device subsidies and connectivity infrastructure as the requisite "entry ticket" for digital citizenship. This protocol is universally applicable to developing economies (e.g., Sub-Saharan Africa, the Caribbean) where soft-skills training often fails due to the lack of physical access to practice those skills.

*5.2.2. Precision Targeting via Persona-Based Service Delivery*

The identification of young women in informal caregiving roles as the highest-leverage segment supports a shift from broad "gender mainstreaming" to persona-based targeting. government information systems should utilize demographic data to segment users not just by income, but by life-stage responsibility. For the "Caregiver" persona, DFL modules should be delivered via mobile-first micro-learning platforms that fit around household duties, rather than scheduled classroom training. This approach maximizes Public Value by directing resources to those with the highest "marginal propensity to learn," as identified by the predictive model's coefficients.

*5.2.3. Algorithmic Filtering for Fiscal Responsibility*

A critical implication for fiscal policy is the management of "deadweight loss." The simulation revealed a decline or stagnation among high-baseline urban professionals, indicating that universal subsidies for this group would be wasteful. Adaptive governance systems must function as exclusionary filters as well as inclusionary ones. By identifying "competency ceilings" *ex-ante*, policymakers can means-test digital subsidies to exclude the digitally proficient, redirecting those funds to the marginalized segments identified above. This "Algorithmic Filtering" provides a defensible, data-driven justification for budget reallocation, essential for maintaining public trust in fiscal decisions.

*5.2.4. Policy Translation: A Conceptual Blueprint for government information systems Integration*

Finally, the proposed framework institutionalizes the adaptive governance theory discussed in Section 2.3 by providing a validated analytical framework. While this study focused on the analytical validation of the process, executing the descriptive profiling, model selection, and scenario simulation offline, the workflow is designed to be conceptually compatible with government information systems.

The study's four-stage pipeline functions as the "intelligence layer" for future digital platforms:
1. Input Standardization: The descriptive profiling (Phase 1) identified the specific variables (e.g., Device Ownership, Digital Spending Tracking) that must be captured by national data registries to establish a baseline.
2. Processing Logic: The selection of the Linear Regression model (Phase 2) provides a transparent, low-latency algorithm suitable for embedding into public sector IT infrastructure, avoiding the "black box" issues of complex neural networks.
3. Decision Support Output: The scenario simulations (Phase 3) demonstrate the specific type of *ex-ante* intelligence, such as projected DFL gains per region, that a government information system needs to display to policymakers for resource prioritization.
4. Adaptive Feedback Logic: Finally, the architecture is designed to support iterative refinement. Conceptually, as new survey data is collected post-intervention, the system would compare realized DFL outcomes against the *ex-ante* projections. Any statistical discrepancies would serve as calibration signals, prompting a retraining of the processing logic. This logic ensures that the system's understanding of citizen behavior evolves over time, preventing the predictive model from becoming static or obsolete as the digital landscape matures.

Currently, policy decisions in the region rely on retrospective reports. By proving that static scenario modeling based on cross-sectional data can yield rigorous prioritization insights (e.g., identifying the 10.1% non-responder cohort), this research offers a proof of value for investing in real-time dashboards. It suggests that if governments integrate this predictive logic into their existing information systems, they can transition from reactive welfare distribution to anticipatory resource allocation. Future research should focus on the technical engineering required to automate this pipeline, turning the study's analytical findings into a live operational tool. It empowers governments to shift from one-size-fits-all provision toward behaviorally informed strategies that are efficient, equitable, and responsive to citizens' lived realities.

*5.3. Theoretical and Practical Contributions*

This study advances the scholarly and practical discourse on digital financial inclusion by integrating predictive analytics with policy simulation in the context of resource-constrained nations. It introduces a framework that not only diagnoses the structural roots of the digital financial divide but also forecasts the effects of targeted interventions. The study contributes to both theoretical advancement and applied innovation in public-sector digital transformation, especially in low-resource settings.

*5.3.1. Theoretical Contributions*

Theoretically, this study advances the discourse on adaptive governance by providing the necessary computational mechanism to support it. Contributing to the Information Systems (IS) literature, the study presents a framework designed to address a critical challenge: optimizing resource allocation within data-fragmented, resource-constrained environments. Unlike data-rich governance models that rely on continuous administrative streams (e.g., real-time transaction logs), this framework demonstrates how periodic, cross-sectional survey data can be repurposed into actionable policy intelligence. This offers a practical pathway for governments to bridge the gap between abstract governance theories, such as collective ambidexterity and sustainable governance and the reality of policy implementation (Djatmiko et al., 2025; Hietala & Päivärinta, 2025). By demonstrating how machine learning can be applied to standard survey data, the study contributes to the theory of the Digital Bureaucracy, showing how data-driven feedback loops can replace static administrative procedures.

Furthermore, the study extends predictive modeling in financial literacy by building on Zhu (2025), demonstrating that supervised machine learning can go beyond identification to simulate population-level outcomes. It deepens the integration of behavioral variables into policy analytics, providing empirical support for the taxonomy of digital vulnerability (Pérez-Escolar & Canet, 2023). The identification of the "caregiver paradox", where young women are economically excluded yet digitally ready, provides empirical grounding for theories of digital vulnerability, moving them from static categories to dynamic, modifiable policy targets.

Finally, the research aligns Explainable AI (XAI) with public trust dynamics. By prioritizing a transparent Linear Regression model over black-box algorithms, the study addresses the transparency concerns raised by previous AI for policy design studies (de Bruijn et al., 2022; van Noordt & Misuraca, 2022). It posits that in the public sector, interpretability is a theoretical prerequisite for legitimacy, reinforcing that model complexity should not supersede administrative accountability.

*5.4.2. Practical Contributions*

Practically, the study delivers a replicable three-stage simulation pipeline: descriptive profiling, machine learning prediction, and behavioral scenario modelling, tailored to decision-makers in low-capacity, data-constrained environments. This pipeline extends the related funding allocation model, enabling geospatial targeting of digital inclusion programs by identifying linguistic and demographic hotspots (Caravaggio et al., 2025).

Crucially, the study validates that simple, interpretable models can achieve high predictive power, encouraging public sector agencies to adopt analytics without over-investing in unexplainable AI. This finding offers a pragmatic path forward for governments in developing nations to leapfrog into data-driven governance by repurposing existing periodic instruments, such as national censuses, household income and expenditure surveys, or financial inclusion demand side surveys, rather than waiting for expensive real-time infrastructure. While this approach requires a minimum investment in data harmonization, specifically ensuring that national surveys capture granular microdata on both digital access and financial usage, it obviates the need for continuous telemetric data streams.

The simulation results also establish a hierarchy of policy levers, supporting a "Digital-First" approach. The finding that improving device ownership alone yields a 4.7% DFL increase reinforces Liang et al. (2025) on the primacy of accessibility. The study's computational framework supports precision policymaking, preventing the misallocation of resources to high-baseline populations and facilitating fiscal efficiency in pursuit of digital equity.

*5.5. Limitations and Future Research*

This study's scope is defined by certain methodological boundaries that point toward productive avenues for future research inquiry.

*5.5.1. Limitations*

First, the study relies on cross-sectional data from 2022. This dataset was selected as the designated empirical baseline for the 2025 UNCDF Pacific Digital Economy Program, establishing the benchmark for the region's financial inclusion goals (United Nations Development Programme, 2025). While this facilitates strategic alignment, the reliance on a static snapshot introduces temporal lag; the simulation assumes statistical drivers identified in 2022 remain stable, potentially overlooking recent infrastructural upgrades or social diffusion effects. Consequently, forecasted gains should be interpreted as baseline potential rather than real-time predictions.

Second, as an *ex-ante* policy tool, the simulation relies on the synthetic manipulation of variables. Real-world implementation may encounter unmeasured "last-mile" barriers, such as behavioral resistance or logistical friction, which the current model cannot fully capture without field validation. Finally, while the framework is transferable, the specific behavioral weights (e.g., the high impact of device ownership) reflect the Pacific context. Application to non-archipelagic geographies like Sub-Saharan Africa would require recalibration to account for local cultural and infrastructural variance.

*5.5.2. Future Research*

To advance the field of digital governance, future research should prioritize methodological advancement. This includes assessing predictive outcomes through Randomized Controlled Trials (RCTs) to close the loop in the Adaptive Governance cycle. Additionally, future iterations should adopt multilevel (mixed-effects) modeling to explicitly capture the hierarchical clustering of citizens within nations, thereby refining the understanding of how national policies interact with local community dynamics.

Furthermore, there is a critical opportunity to translate this analytical framework into operational government information systems. Future scholarship should focus on integrating Explainable AI (XAI) dashboards that visualize these simulations for non-technical policymakers (de Bruijn et al., 2022). By developing longitudinal panel datasets to feed these dashboards, governments can adapt the current model from a one-time planning tool into a longitudinal monitoring capable of real-time policy adjustment.

## 6. Conclusion

This study addresses the critical challenge of financial exclusion by introducing a predictive simulation framework that supports adaptive governance within resource-constrained environments. By integrating descriptive analytics with transparent machine learning, the research transforms static survey data into dynamic policy intelligence. This approach allows government information systems to shift from retrospective monitoring to anticipatory decision-making, facilitating digital transformation strategies that are not merely reactive but are evidence-based, equity-driven, and fiscally responsible.

The empirical analysis highlights that foundational digital capabilities and active behavioral patterns, rather than demographic attributes alone, are the primary drivers of the digital financial divide. The simulation results challenge the efficacy of broad-spectrum educational campaigns, supporting a "digital-first" sequencing protocol where infrastructure access serves as the requisite baseline for behavioral interventions. Furthermore, the ability to identify high-leverage segments alongside "non-improvers" demonstrates the utility of algorithmic filtering in public administration. This capability enables governments to optimize resource allocation by directing funds away from the already proficient and toward the most responsive excluded groups, thereby preventing policy redundancy and maximizing social return on investment.

Crucially, while grounded in the Pacific context, this computational workflow offers a scalable blueprint for digital governance globally. The framework addresses a universal administrative challenge: optimizing decision-making under conditions of uncertainty and data fragmentation. Consequently, this approach is highly transferable to other archipelagic or developing regions seeking to leapfrog into data-driven policymaking without the immediate need for expensive real-time infrastructure. Ultimately, this research bridges the gap between technical possibility and administrative reality, providing a replicable mechanism to dismantle structural barriers and enable the digital state that is designed for its most vulnerable citizens.

**Appendix A. Descriptive Statistics of Digital Financial Competency (DFC) and Total Literacy (DFL) by Country**

| Country | DFC | | | | | DFL | | | | |
|---|---|---|---|---|---|---|---|---|---|---|
| | **Mean** | Median | Std | Min | Max | **Mean** | Median | Std | Min | Max |
| **Fiji** | **43.7%** | 44.4% | 16.4% | 0.0% | 100.0% | **50.7%** | 51.9% | 14.8% | 11.5% | 92.3% |
| **PNG** | **32.6%** | 33.3% | 15.0% | 0.0% | 77.8% | **41.1%** | 38.5% | 14.3% | 3.8% | 88.5% |
| **Samoa** | **38.7%** | 33.3% | 15.7% | 0.0% | 77.8% | **43.3%** | 42.3% | 12.4% | 0.0% | 86.5% |
| **Solomon Islands** | **36.3%** | 33.3% | 11.9% | 0.0% | 77.8% | **41.9%** | 40.4% | 12.2% | 7.7% | 82.7% |
| **Timor-Leste** | **37.9%** | 33.3% | 13.0% | 0.0% | 88.9% | **39.6%** | 38.5% | 12.7% | 3.8% | 84.6% |
| **Tonga** | **46.2%** | 44.4% | 15.2% | 0.0% | 88.9% | **44.3%** | 44.2% | 12.5% | 7.7% | 84.6% |
| **Vanuatu** | **37.3%** | 44.4% | 17.6% | 0.0% | 88.9% | **44.2%** | 44.2% | 12.9% | 1.9% | 76.9% |